\documentstyle[12pt]{article}

\renewcommand{\thefootnote}{\fnsymbol{footnote}}
\newcommand{\r}[1]{(\ref{#1})}
\begin{document}
\thispagestyle{empty}
\newlength{\defaultparindent}
\setlength{\defaultparindent}{\parindent}

\begin{center}
{\large{\bf A HIERARCHY OF GAUGED GRASSMANIAN MODELS IN $4p$
DIMENSIONS WITH SELF-DUAL INSTANTONS}} \vspace{1.5cm}\\

{\large R.P. Manvelyan\footnote{E-MAIL:manvelyan@vxc.yerphi.am
},}\vspace{0.5cm}

{\it Theoretical Physics Department,}\\
{\it Yerevan Physics Institute}\\
{\it Alikhanyan Br. st.2, Yerevan, 375036 Armenia }\vspace{1.5cm}\\

{\large D.H. Tchrakian\footnote{E-
MAIL:tchrakian@vax1.may.ie},}\vspace{0.5cm}

{\it Department of Mathematical Physics,}\\
{\it St Patrick's College Maynooth,}\\
{\it Maynooth, Ireland}\\
{\it School of Theoretical Physics,}\\
{\it Dublin Institute for Advanced Studies,}\\
{\it 10 Burlington Road,}\\
{\it Dublin 4, Ireland}

\end{center}
\bigskip
\bigskip
\bigskip
\bigskip
\begin{abstract}We present a hierarchy of gauged Grassmanian models in
$4p$ dimensions, where the gauge field takes its values in the $2^{2p-
1}\times 2^{2p-1}$ chiral representation of SO(4p). The actions of all these
models are absolutely minimised by a hierarchy of self-duality equations,
all of which reduce to a single pair of coupled ordinary differential
equations when subjected to $4p$ dimensional spherical symmetry.
\end{abstract}

\vfill
\setcounter{page}0
\renewcommand{\thefootnote}{\arabic{footnote}}
\setcounter{footnote}0
\newpage

\newcommand{\ra}{\rightarrow}

\pagestyle{plain}

\section{\bf Gauging Sigma Models}
\setcounter{equation}{0}

\parskip0.5truecm

In addition to introducing covariant derivatives in terms of the gauge
connection, and curvature dependent terms that describe the dynamics of
the gauge field, our prescription for gauging a Sigma model includes a
crucial additional requirement: the action/energy density of the gauged
Sigma model must be bounded from below by a total divergence whose
(surface) integral can be arranged to take on a non vanishing value by
requiring suitable asymptotic conditions for the classical solutions. These
are the topologically stable instantons/solitons with finite action/energy.

The gauging of Sigma models in 3 dimensions has been considered by
Witten\cite{Witten} and Rubakov\cite{Rubakov} for the Skyrme
model\cite{Skyrme}, which is equivalent to the $O(4)$ Sigma model. In both
these cases there is no topological inequality bounding the energy from
below, and indeed the lack of (topological) stability has been
exploited\cite{Rubakov} in the context of Techniskyrmions.

By contrast, the gauging of the $O(3)$ Sigma model in 2 dimensions was
considered by Mehta, Davis and Aitchison\cite{Aitchison} in the context of
a 2+1 dimensional Chern-Simons theory, where finite energy vortex
solutions were found. The topological inequalities that are responsible for
the stability of these vortices were given in Ref\cite{Piette} and
subsequently this analysis was extended\cite{Tchrakian1} to an
$O(3)$model augmented by a Skyrme term. The topological inequalities
employed in Refs.\cite{Aitchison},\cite{Piette},\cite{Tchrakian1} were
stated not for the $O(3)$ model but for the classically equivalent $CP^{1}$
model which happens to be a Grassmanian model. In Ref.\cite{Tchrakian1},
it was described in some detail how the gauging of the $O(3)$ model with a
$U(1)$ gauge field did not lead to a non-trivial topological lower bound on
the energy while gauging the equivalent $CP^{1}$ model did just that.

Motivated by the above, instead of gauging $2n$ dimensional $O(2n+1)$
Sigma models for which we do not expect to find topological inequalities, we
consider the gauging of a certain hierarchy of Grassmanian models in
$2n$ dimensions which we shall show to be endowed with the requisite
topological inequalities implying stability and lower bounds on their
actions. These $2n$ dimensional Grassmanian models will be gauged with a
connection field taking its values in the chiral representation of $SO(2n)$.
If we allowed the integer $n$ to be both even and odd, then this would have
been a direct generalistaion of the $SO(2) \approx U(1)$ gauged $2$
dimensional $CP^{1}$ model of
Refs.\cite{Aitchison},\cite{Piette},\cite{Tchrakian1}, where the $U(1)$
gauged $CP^{1}$ model would have constituted the first member of the $2n$
dimensional hierarchy of $SO(2n)$ gauged Grassmannian models. In this
paper we restrict ourselves to only even values of the integer $n$ because
we intend to restrict ourselves to a hierarchy of scale invariant systems. A
more general classification including systems that are not scale invariant
will be undertaken elsewhere. As a consequence of their scale invariance
the Euler-Lagrange equations of this hierarchy of models are solved by a
hierarchy of first order self-duality equations. The self-duality equations
are presented, and this is followed by some discussion of the physical
relevance of the $4$ dimensional case.

\section{\bf The Gauged Grassmanian Models}

The Grassmanian models in $2n$ dimensions that we consider are described
by the fields
\begin{equation}
\label{1}
z^A{}_i =(z^a{}_i ,z^{\alpha}{}_i ),\       i,a,\alpha=1,...,2^{n-1};
\end{equation}
subject to the constraint
\begin{equation}
\label{2}
(z^*)^i{}_A z^A{}_j =\delta^i{}_j
\end{equation}
or more briefly $z^\dagger z=1$. The covariant derivative of the field $z$ is
defined by
\begin{equation}\label{3}
D_{\mu}z=\partial_{\mu}z-zA_{\mu}
\end{equation}
where the connection $A_{\mu}$ takes its values in the $2^{n-1} \times
2^{n-1}$ dimensional chiral representation of $SO(2n)$, and is not here the
composite field $A_{\mu}=z^\dagger \partial_{\mu}z$ of a pure
Grassmanian model.

The action of the local gauge group element $g$ is given by
\begin{equation}\label{4}
z\rightarrow zg
\end{equation}
\begin{equation}\label{5}
D_{\mu}z\rightarrow D_{\mu}zg
\end{equation}

Let us henceforth restrict to even values of $n=2p$ and define the
following three tensor field strengths $F(2p)$, $G(2)$ and $H(2p)$ as
follows: The $2p$ form field $F(2p)$ is the $p$ fold totally antisymmetrised
product of the curvature $2$ form $F(2) \equiv F_{\mu\nu}$
\begin{equation}
\label{6}
F(2p) \equiv
F_{\mu_{1}\mu_{2}...\mu_{2p}}=F_{[\mu_{1}\mu_{2}}...F_{\mu_{2p-
1}\mu_{2p]}}
\end {equation}
The square brackets on the indices $\mu_{1},\mu_{2},..$ imply total
antisymmetrisation. The $2$ form field $G(2)$ is defined by
\begin{equation}
\label{7}
G(2) \equiv G_{\mu\nu}=D_{[\mu}z^\dagger D_{\nu]}z
\end{equation}
The $2p$ form field $H(2p)=G(2) \wedge F(2p-2)$ is then given by
\begin{equation}
\label{8}
H(2p) \equiv
H_{\mu_{1}\mu_{2}...\mu_{2p}}=G_{[\mu_{1}\mu_{2}}F_{\mu_{3}...\mu_{2
p}]}
\end{equation}
Note that $G(2)$ and hence also $H(2p)$ are both gauge covariant
quantities as seen from \r{5}, since $F(2p)$ is by definition gauge
covariant. The curvature $2$ form here is defined by $D_{[\mu}D_{\nu]}z=-
zF_{\mu\nu}$. Note also that $H(2p)$, which depends on the curvature
form $F(2p-2)$ is independent of the curvature $2$ form $F(2)$ in $4$
dimensions only, in which case $H(2)=G(2) \wedge F(0) \equiv G(2)$.

Using the $F$ and $H$ field strengths, we proceed to state the inequalities
which will give rise to topological stability, and at the same time define the
action densities which are bounded from below by the respective
topological charge densities. These inequalities are
\begin{equation}
\label{9}
Tr[ ^{\star} F(2p) \mp H(2p)]^2 \geq 0
\end{equation}
where the notation $^{\star}F$ implies the definition for the Hodge dual of
$F$. The inequality \r{9} yields both the definition of the action density
and the topological inequality
\begin{equation}
\label{10}
{\cal L} _{p} \stackrel{\rm def}{=}Tr[F(2p)^2 +H(2p)^2] \geq \pm 2 Tr
^{\star} F(2p)H(2p)
\end{equation}
In \r{9} and \r{10}, the trace is taken over the indices of the local gauge
group, c.f. \r{5}, and hence all quantities are gauge invariant.

It is easy to verify that the right hand side of \r{10} is a total divergence.
This is seen if we reexpress this density as $Tr^{\star}F(4p-2)G(2)$, and
make use of the Bianchi identities as well as the constraint (2). The
topological charge density $\varrho$ is then
\[\varrho =Tr(^{\star}F(4p-2))_{\mu\nu}G_{\mu\nu}\]
\begin{equation}
\label{11}
\simeq \varepsilon
_{\mu_{1}\mu_{2}\mu_{3}\mu_{4}...\mu_{4p}}[TrF_{\mu_{1}\mu_{2}}F_{\
mu_{3}\mu_{4}}...F_{\mu_{4p-1}\mu_{4p}}+2\partial
_{\mu_{1}}TrF_{\mu_{3}\mu_{4}}...F_{\mu_{4p-
1}\mu_{4p}}z^{\dagger}D_{\mu_{2}}z]
\end{equation}

Thus the hierarchy of gauged Grassmanian models in every $4p$
dimension is characterised by the Lagrangian density ${\cal L}_{p}$ on the
left hand side of \r{10}, which is bounded from below by the total
divergence \r{11}. We notice that the first term in \r{11} is the $2p$-th
Chern-Pontryagin(C-P) density. This quantity is the total divergence of the
Chern-Simons(C-S) density while the second term is already in total
divergence form. That solutions exist for which these surface integrals are
finite and nonvanishing will be shown below when we restrict to the
spherically symmetric field configurations.

We notice here that the $4p$ dimensional action integral of the Lagrange
density ${\cal L}_{p}$ given by the left hand side of \r{10} is scale
invariant. This is because the $2$ form field $G_{\mu\nu}$ defined in \r{7}
scales exactly like the curvature $2$ form $F_{\mu\nu}$ since the
Grassman-valued fields $z$ are themselves dimensionless. This is reflected
by the absence of a dimensional constant in the inequality \r{9}, which
means that the self duality equations which saturate \r{9}
\begin{equation}
\label{12}
^{\star}F(2p)=\pm H(2p)
\end{equation}
also do not feature a dimensional constant.

The close analogy with the $4p$ dimensional Yang-Mills
hierarchy\cite{Tchrakian2} \cite{Tchrakian3} suggests that the relevant
solutions of \r{12} will have a pure-gauge behaviour at infinity. For the
latter statement, it is necessary to first choose the gauge group, and since
equations \r{12} are not restricted to take values inside the algebra of this
gauge gruop, we must specify further the representation in which the
connection field is defined. This was stated above as being the chiral
representation of $SO(4p)$, for the $4p$ dimensional member of this
hierarchy.

\section{\bf Spherically Symmetric Fields}

To explain the problem of imposing spherical symmetry, we find it most
convenient to employ the formalism of Schwarts\cite{Schwarts}. Very
briefly, this amounts to solving an algebraic constraint for each of the two
fields involved, namely the gauge connection $A_{\mu}=(A_{i},A_{4p})$,
$i=1,...,4p-1$ and the Grassmanian field $z$. These equations stated at the
north pole of the $4p$-sphere, are, for the gauge connection
\begin{equation}
\label{13}
A_{0}=0
\end{equation}
\begin{equation}
\label{14}
\mu ^{-1}A_{i}\mu =h_{i}{}^j A_{j}
\end{equation}
where $h$ is an element of $SO(4p-1)$, the stability group of the north
pole, and $\mu$ is the appropriate representation of $h$ which in our case
is the chiral representation of $SO(4p)$ in terms of gamma matrices. For
the Grassmanian field the spherical symmetry constraint is
\begin{equation}
\label{15}
z\mu=z
\end{equation}
While equation\r{14} can be solved readily as was done in
Ref.\cite{Tchrakian3} and elsewhere, equation\r{15} cannot be solved
except for the case $n=1$ in equation\r{1}, namely the case of the $CP^1$
model with $U(1)$ gauge freedom in $2$ dimensions which was considered
in Ref\cite{Aitchison}. This is not surprising since in $2$ dimensions the
$CP^1$ model is equivalent to the $O(3)$ model, and the symmetry equation
corresponding to \r{15} for all $d$ dimensional $O(d+1)$
models\cite{Dolan} is solvable. This means that any radially symmetric
Ansatz that we make here will not be strictly spherically symmetric and
hence due care must be taken for its consistency. This involves the
consistency of the Euler-Lagrange equations arising from the variation of
the one dimensional subsystem resulting from the application of the
Ansatz, with the Euler-Lagrange equations of the system obtained before
the appliaction of the said Ansatz.

In the present paper, the question of this consistency of the Ansatz does
not arise because we shall not be applying the variational principle to the
reduced one dimensional subsystem, but instead will be solving the full self
duality equations. In future applications however, when we would envisage
extending the present models \r{10} by higher/lower order terms in the
derivatives of the fields, the solutions would be necessarily non-self dual
and hence it would be necessary to verify the consistency of the Ansatz.

Our Ansatz for the connection $A_{\mu}$ satisfies the symmetry restriction
\r{14} and is therefore strictly spherically symmetric. Using $r$ as the
radial variable and $\hat x_{\mu}$ as the unit vector in $4p$ dimensions,
this is
\begin{equation}
\label{16}
A_{\mu}=\frac{2}{r} [1-k(r)]\tilde \Sigma _{\mu\nu}\hat x_{\nu}
\end{equation}
The tensor valued matrices $\tilde \Sigma_{\mu\nu}$ in \r{16}, and their
chiral conjugates $\Sigma_{\mu\nu}$ are defined by
\begin{equation}
\label{17}
\tilde \Sigma_{\mu\nu}=-\frac{1}{4}\tilde \Sigma_{[\mu}\Sigma_{\nu]}
,\qquad \Sigma_{\mu\nu}=-\frac{1}{4}\Sigma_{[\mu}\tilde \Sigma_{\nu]}
\end{equation}
in terms of the vector valued matrices $\tilde \Sigma_{\mu}$ and
$\Sigma_{\mu}$ are in turn defined in terms of the gamma matrices
$\Gamma_{\mu}$ in $4p$ dimensions and the respective chiral matrix
$\Gamma_{4p+1}$ as
\begin{equation}
\label{18}
\tilde \Sigma_{\mu}=\frac{1}{2}(1-\Gamma_{4p+1})\Gamma_{\mu},\qquad
\Sigma_{\mu}=\frac{1}{2}(1+\Gamma_{4p+1})\Gamma_{\mu}
\end{equation}
Our Ansatz for the Grassmanian valued field
$z^A{}_{i}=(z^a{}_{i},z^{\alpha}{}_{i})$ is
\begin{equation}
\label{19}
z^a{}_{i}=sin\frac{f(r)}{2}\,\delta ^a{}_{i},\qquad
z^{\alpha}{}_{i}=cos\frac{f(r)}{2}\,\hat
x_{\mu}(\Sigma_{\mu})^{\alpha}{}_{i}
\end{equation}

It is now a straightforward matter to compute the curvature $2$ form
$F_{\mu\nu}$ and the covariant derivative $D_{\mu}z$ for the field
configuration given by \r{16} and \r{19} and hence to evaluate the $2p$
form fields $F(2p)$ and $H(2p)$, with a view to substituting the latter in the
self duality equations\r{12}. To analyse the resulting equations in the
functions $k(r)$ and $f(r)$ we shall need the tensor-spinor identities
employed in Ref.\cite{Tchrakian3}
\begin{equation}
\label{20}
\Sigma (2p)=^{\star}\Sigma (2p), \qquad \tilde \Sigma (2p)=-^{\star}\tilde
\Sigma (2p)
\end{equation}
where we have used the following notation for $\Sigma(2p)\equiv
\Sigma_{\mu_{1}\mu_{2}...\mu_{2p}}$ and $\tilde \Sigma(2p)\equiv \tilde
\Sigma_{\mu_{1}\mu_{2}...\mu_{2p}}$, analogous with \r{6},
\begin{equation}
\label{21}
\Sigma_{\mu_{1}\mu_{2}...\mu_{2p}}=\Sigma_{[\mu_{1}} \tilde
\Sigma_{\mu_{2}}...\Sigma_{\mu_{2p-1}}\tilde \Sigma_{\mu_{2p}]},\quad
\tilde \Sigma_{\mu_{1}\mu_{2}...\mu_{2p}}=\tilde
\Sigma_{[\mu_{1}}\Sigma_{\mu_{2}}...\tilde \Sigma_{\mu_{2p-
1}}\Sigma_{\mu_{2p}]}
\end{equation}
The resulting first order differential equations for the functions $k(r)$
and $f(r)$ are the same for all $p$, namely for all members of the
hierarchy. This is exactly the same situation as for the $4p$ dimensional
hierarchy of scale invariant Yang-Mills\cite{Tchrakian2} models. In that
case\r{Tchrakian3} the corresponding hierarchy of self duality equations
reduced to a single first order equation, which yields the well known BPST
instanton\cite{BPST}.

In the present case, the these coupled first order equations are, say in the
self dual case,
\begin{equation}
\label{22}
k'=\frac{2}{r}[k^2+\frac{1}{2}(2k-1)(cosf-1)],\quad \qquad f'sinf=-
\frac{1}{r}k(k-1)
\end{equation}
It should be noted here that the radial variable $r$ in \r{22} can be
replaced with the dimensionless variable $\rho $, where $\rho =\lambda r$
in terms of an arbitrary scale $\lambda $. As a result, these instantons will
be localised to an arbitrary scale, which is a consequence of the scale
invariance of ${\cal L}_{p}$.

The asymptotic behaviour of the solutions will be found in the next section
where we examine the $4$ dimensional $p=1$ case in detail. We shall see
there, that the asymptotics found for the $p=1$ case hold also for all $p$. We
shall therefore anticipate this large $r$ behaviour and state that for
arbitrary $p$ the solutions of \r{22} will be expressed in terms of the
matrix $\omega =\hat x_{\mu}\tilde \Sigma_{\mu}$ in the pure-gauge
form
\begin{equation}
\label{23}
A_{\mu}=\omega ^{-1}\partial _{\mu}\omega,\qquad z^a{}_{i}=0,\qquad
z^{\alpha}{}_{i}=\omega^{\alpha}{}_{i}
\end{equation}
which leads to the usual instanton charge when substituted in the
topological charge density $\rho$ given by \r{11}. \r{23} is a consequence
of the scale invariance of the hierarchy of gauged Grassmanian models
characterised by the Lagrangians ${\cal L} _{p}$ given in \r{10}. The
pure-gauge behaviour \r{23} can also be understood when we note that
otherwise, if $F_{\mu\nu}$ decayed as the inverse square of the radius as it
does in the hierarchy of $SO(d)$ Higgs models in $d$
dimensions\cite{Tchrakian4} , the square of the $2p$ form field $F(2p)$ in
${\cal L}_{p}$ here would cause the logarithmic divergence of the action.

Before proceeding to the details of $p=1$, let us note two features of the
general $p\geq 1$ case. The first is that only in the $p=1$ case is the matrix
$\omega $ an element of the full gauge group, chiral $SO(4) \equiv
SO(4)_{\pm }$, namely $SU(2)$. In all other cases with $p>1$, $\omega $ is
parametrised by the $4p-1$ independent components of $\hat x_{\mu}$
which is successively smaller than the number of parameters of $SO(4p)$,
with increasing $p$. The second feature is that the self duality equations
\r{12} are successively more overdetermined for increasing $p>1$. This is
the case also for the Yang-Mills hierarchy of self duality
equations\cite{Tchrakian5}, in which case for $p>1$ the least symmetric
solution is the axially symmetric instanton\cite{Tchrakian6}. Here
counting the number of equations in \r{12} is equal to
$\frac{(4p)!}{(2p)!^2}$. This number increases with $p$ while the number
of fields does not. Clearly the imposition of full spherical symmetry
resulted in equations \r{22} which are not overdetermined. The question as
to the existence of axially symmetric or other solutions belongs to the
subject of future investigations.

\section{\bf $p$=1: Four Dimensions}

We consider this case on its own in more detail not only because this is the
physically interesting case, but also as the simplest member of the
hierarchy it has some special properties. The Lagrange density of the $p=1$
member of the hierarchy is
\begin{equation}
\label{24}
{\cal L}_{1}=Tr[F_{\mu\nu}{}^2+(D_{[\mu}z^{\dagger}D_{\nu]})^2]
\end{equation}
and the topological charge density is
\begin{equation}
\label{25}
\varrho =4\varepsilon
_{\mu\nu\rho\sigma}TrF_{\mu\nu}D_{\mu}z^{\dagger}D_{\nu}z
\end{equation}

Substituting the field configuration given by \r{16} and \r{19} into \r{24}
and \r{25} and integrating over the angular volume, we find the one
dimensional effective Lagrange density $L$ and topological charge density
$\sigma$ descending from \r{24} and \r{25} repectively to be
\begin{equation}
\label{26}
L=rk'^2+\frac{4}{r}k^2 (k-1)^2 +rf'^2 sin^2 f+\frac{4}{r}[k^2
+\frac{1}{2}(2k-1)(cosf-1)]^2
\end{equation}
\begin{equation}
\label{27}
\sigma=\frac{d}{dr}[\frac{1}{3}k^3 +\frac{1}{2}k(k-1)(cosf-1)]
\end{equation}
Note that the charge density $\sigma$ for the spherically symmetric field
configuration is a total derivative \r{27}.

By inspecting the Lagrangian $L$ \r{26} carefully, we conclude that there
is only one set of asymptotic values for the functions $k(r)$ and $f(r)$
which is consistent with convergent action integral and non singular
behaviour at the origin of $r$. This is
\begin{equation}
\label{28}
\lim_{r \rightarrow 0} k(r)=1 \qquad \qquad  \lim_{r \rightarrow \infty}
k(r)=0
\end{equation}
\begin{equation}
\label{29}
\lim_{r \rightarrow 0} f(r)= \pi  \qquad \qquad  \lim_{r \rightarrow
\infty} f(r)=0
\end{equation}

The asymptotic values \r{28} and \r{29} were deduced above by examining
the reduced Lagrange density \r{26} for $p=1$. It is not hard however to
verify that \r{28} and \r{29} remain valid in the $p>1$ case as well. This
justifies our anticipation of the pure-gauge behaviour \r{23} above, which
is simply a consequence of the second member of \r{28}.

Another interesting feature of the our $4p$ dimensional instantons can be
learnt from the $p=1$ case. This is the dependence of the topological charge
on the small $r$ behaviour of the function $f(r)$ parametrising the
Grassmanian field $z$. One might expect that the charge of our spherically
symmetric instanton, regarded as the Hedgehog of a Grassmanian Sigma-
Skyrme model, should depend on the small $r$ behaviour. In the first
member of \r{29} $\pi $ can obviously be replaced by $N\pi $ for any
integer $N$. When this is done for the usual $3$ dimensional Skyrme model
\cite{Skyrme} it is found that its topological charge (Baryon number) takes
the value $N$ and its energy becomes 2.98 times the energy of the  unit
charge Hedgehog\cite{Jackson}. It is interesting to see whether changing
$\pi $ to $N\pi $ in the first member of \r{29} results in any change of the
topological charge? The answer follows immediately from an inspection of
\r{27}. It is clear that the integral of \r{27} between the limits given by
\r{28} and \r{29} is insensitive to this change and hence that our
spherically symmetric instantons have a unique topological charge of one
unit.

Interpreting the instanton of the model \r{25} as the vacuum\cite{Jackiw}
is quite natural since it behaves as a pure gauge at large $r$. On the other
hand because of the scale invariance of\r{25} the instanton is localised to
an arbitrary scale. If we wish to have an instanton localised to an absolute
scale, we would have to add to the Lagrangian \r{24} several lower/higher
order terms in the derivatives of the fields. These would be the quadratic
kinetic term $TrD_{\mu}z^{\dagger}D_{\mu}z$ and possibly a potential
term. Both of these scale with a smaller power than \r{24} does, and hence
to satisfy Derrick's theorem a higher order Skyrme term such as a sextic or
an octic term in the derivatives must also be added to \r{24}. This is a task
of
some complexity and is deferred to a future investigation.

Having mentioned the question of the vacuum of the theory being
described by the instanton of the model, we end with the corresponding
question pertaining to the finite energy solutions in the static limit. This is
the question analogous to the one dealt with by Rubakov\cite{Rubakov} in
the case of the $SU(2)$ gauged Skyrme model. As in Ref.\cite{Rubakov} we
would not expect to find a stable finite energy "soliton" solution. Indeed,
the simplest way to demonstrate this is to proceed in the usual way by
calculating the static Hamiltonian of the model in the temporal gauge and
studying the finite energy solutions. Because of the multiplet structure of
the Grassmanian field $z$, it is not possible to subject this field to
spherical
symmetry in $3$ dimensions. This is completely analogous to the case in the
Weinberg-Salam model\cite{Manton} and the $SO(4)\times U(1)$ Higgs
model\cite{Tchrakian7} where the solution is the unstable sphaleron
characterised by the non contractible loop(NCL) field configuration for the
field $z^A{}_{i}=(z^a{}_{i},z^{\alpha}{}_{i})$
\begin{equation}
\label{30}
z^a{}_{i}=cos\frac{\tilde f}{2} \,\delta^a{}_{i},\qquad
z^{\alpha}{}_{i}=sin\frac{\tilde
f}{2}\,p_{\mu}(\mu,\theta,\phi)(\Sigma_{\mu})^{\alpha}{}_{i}
\end{equation}
where $\Sigma_{\mu}=\tau_{\mu}=(\tau_{i},\tau_{4})=(i\sigma_{i},1)$ in
the familiar $p=1$ case and $\tilde f(\tilde r)$ is a function of the radial
variable $\tilde r$ in $3$ dimensions, and the $4$ vector
$p_{\mu}(\mu,\theta,\phi)$ is
\begin{equation}
\label{31}
p_{\mu}=(sin\mu sin\theta cos\phi ,sin\mu sin\theta sin\phi ,
sin^2\mu cos\theta +cos^2\mu,sin\mu cos\mu (cos\theta -1))
\end{equation}
in which $\theta$ and $\phi$ are the polar and azimuthal angles in $3$
dimensions while $\mu$ is a constant parametrising the NCL. The
sphaleron solution corresponds to the value $\mu =\frac{\pi}{2}$ and is
unstable against the variation of this parameter. A complete study of the
sphaleron solutions of this model will be given in detail elsewhere.

{\bf Acknowledgements} This work was suppotred in part by INTAS
contracts 93-1630 and 93-2058. One of us(R.P.M.) acknowledges the
hospitality of the School of Theoretical Physics of the Dublin Institute for
Advanced Studies, where this work was finished.


\newpage

\begin{thebibliography}{99}

\bibitem{Witten} E. Witten, Nucl. Phys. {\bf 223} (1983) 433.
\bibitem{Rubakov} V.A. Rubakov, Nucl. Phys. {\bf 256} (1985) 509.
\bibitem{Skyrme}T.H.R. Skyrme, Proc. Roy. Soc.{\bf A260} (1961)
127;\\Nucl.Phys. {31} (1962) 556.
\bibitem{Aitchison}M. A. Mehta, J.A. Davis and I.J.R. Aitchison,
Phys.Lett.{\bf B281}(1984) 86.
\bibitem{Piette} , B.M.A.G. Piette, D.H. Tchrakian and W.J. Zakrzewski, Phys.
Lett {\bf B339} (1994) 95.
\bibitem{Tchrakian1} "Solitons in Gauged Sigma Models: 2 dimensions",
DIAS-STP-95-16.
\bibitem{Tchrakian2} D.H. Tchrakian, J. Math. Phys.{\bf 21} (1980) 166;
\\D.O'S\'e and D.H. Tchrakian, Lett. Math. Phys.{\bf 13} (1987) 211.
\bibitem{Tchrakian3} D.H. Tchrakian, Phys.Lett.{\bf B150} (1985) 360.
\bibitem{Schwarts} A.S. Schwarts, Commun. Math. Phys. {56} (1977) 79.
\bibitem{Dolan}B.P. Dolan and D.H. Tchrakian, Phys.Lett.{\bf B198}(1987)
447.
\bibitem{BPST}A.A. Belavin, A.M. Polyakov, A.S. Schwarts and Yu.S.
Tyupkin, Phys. Lett. {\bf B59} (1975) 85.
\bibitem{Tchrakian4}Zh-Q. Ma and D.H. Tchrakian, Lett. Math. Phys. {\bf
26} (1992) 179.
\bibitem{Tchrakian5}D.H. Tchrakian and A. Chakrabarti, J. Math. Phys.
{\bf 32} (1991) 2532.
\bibitem{Tchrakian6}A. Chakrabarti, T.N. Sherry and D.H. Tchrakian, Phys.
Lett. {\bf B162} (1985) 340; \\J. Burzlaff, A. Chakrabarti and D.H. Tchrakian,
J. Math. Phys. {\bf 34} (1993)1665.
\bibitem{Jackson}A.D. Jackson and M. Rho, Phys. Rev. Lett. {\bf 51}(1983)
751.
\bibitem{Jackiw}R. Jackiw and C. Rebbi, Phys. Rev. Lett. {\bf 37} (1976) 172.
\bibitem{Manton} N.S. Manton, Phys. rev. {\bf D28} (1983) 2019.
\bibitem{Tchrakian7}G.M. O'Brien and D.H. Tchrakian, Mod. Phys. Lett. {\bf
A4} (1989) 1389; \\Phys. Lett. {\bf 282} (1992) 111.
\end{thebibliography}
\end{document}